\documentclass{PoS}
\usepackage{graphicx}
\usepackage{bm}
\usepackage{multirow}
\usepackage{epsfig}
\usepackage{psfrag}
\usepackage{soul}
\usepackage{multirow}
\usepackage{mathptmx}
\usepackage{mathrsfs}
\usepackage{amsmath, amssymb}
\usepackage{dsfont}
\newcommand{\ident}{\mathds{1}}

\title{Conserved currents for Mobius Domain Wall Fermions}

\ShortTitle{Mobius Conserved Currents}

\author{\speaker{Peter A Boyle}\\
        The University of Edinburgh\\
        RBC and UKQCD collaborations\\
        E-mail: \email{paboyle@ph.ed.ac.uk}}

\abstract{We derive the exactly conserved vector, and almost conserved axial currents for
rational approximations to the overlap operator with a general Mobius kernel. The
approach maintains manifest Hermiticity, and allows matrix elements of the
currents to be constructed at no extra cost after solution of the usual 5d system of equations,
similar to the original approach of Furman and Shamir for domain wall Fermions.
}

\FullConference{The 32nd International Symposium on Lattice Field Theory,\\
		23-28 June, 2014\\
		Columbia University New York, NY}

\begin{document}

RBC and UKQCD have recently adopted the M\"obius  
generalisation of the 
domain wall action~\cite{Brower:2004xi,Brower:2005qw,Brower:2012vk},
Our conventions are as follows. The usual Wilson matrix is 
$D_W(M) = M+4 - \frac{1}{2} D_{\rm hop},$
where
$
D_{\rm hop} = (1-\gamma_\mu) U_\mu(x) \delta_{x+\mu,y} +
              (1+\gamma_\mu) U_\mu^\dagger(y) \delta_{x-\mu,y} \,.
$
We introduce the five dimensional action
$
S^5 = \bar{\psi} D^5_{GDW} \psi
$
where
\begin{eqnarray}
D^5_{GDW}
&=&
\left(
\begin{array}{cccccc}
  \tilde{D} & - P_- & 0& \ldots & 0 &  m P_+ \\
-  P_+  & \ddots &  \ddots & 0      & \ldots &0 \\
0     & \ddots &  \ddots & \ddots & 0      &\vdots \\
\vdots& 0      &  \ddots & \ddots & \ddots & 0\\
0     & \ldots &    0    &  \ddots& \ddots & - P_- \\
m P_- & 0      & \ldots  &  0     &   - P_+   
& \tilde{D}
\end{array}
\right)\,,
\end{eqnarray}
and we define
$
D_+ = (b D_W + 1) \quad;\quad
D_-= (1-c D_W )\quad;\quad
\tilde{D} = (D_-)^{-1} D_+\,.
$
This generalized set of actions reduces to the 
standard Shamir action in the limit $b=1$, $c=0$, and it can also be
taken to give the polar approximation to the 
Neuberger overlap action as another limiting case~\cite{Borici:1999zw,Borici:1999da}. 
In all of our simulations we take the coefficients $b$ and $c$ as constant across the fifth dimension,
As in the Shamir domain wall fermion formulation we identify ``physical'', four-dimensional quark fields $q$ and $\bar{q}$ whose Green's functions define our domain wall fermion approximation to continuum QCD.   We choose to construct these as simple chiral projections of the five-dimensional fields $\psi$ and $\bar{\psi}$ which appear in the action.
\begin{equation}
\begin{array}{cccccc}
q_R = P_+ \psi_{L_s} &\quad& q_L = P_- \psi_{1}\, & \bar{q}_R = \bar{\psi}_{L_s} P_- & \quad & \bar{q}_L =  \bar{\psi}_{1} P_+\,.
\label{eq:physical_fields}
\end{array}
\end{equation}
The choice of physical quark fields given in Eq.~\eqref{eq:physical_fields} has the added benefits that the corresponding four-dimensional propagators satisfy a simple $\gamma^5$ hermiticity relation and a hermitian, partially conserved axial current can be easily defined.
If we introduce the so-called transfer matrix as
\begin{equation}
\begin{array}{ccc}
T^{-1} &=& -(Q_-)^{-1} Q_+ = -[ H_M - 1]^{-1} [ H_M + 1].
\end{array}
\end{equation}
and define the M\"obius kernel as
\begin{equation}
H^M = \gamma_5\frac{(b+c) D_W}{2+(b-c)D_W}\,.
\end{equation}
One can show that $D^5_\chi$ takes the following form\cite{Brower:2004xi,Brower:2005qw,Brower:2012vk},
\begin{eqnarray}
D^5_\chi&=& 
\left[ 
\begin{array}{cccccc}
P_- - m P_+ & -T^{-1} & 0 & \ldots & \ldots & 0\\
0               & 1 & -T^{-1} & 0    & \ldots & \vdots\\
\vdots          & 0   & \ddots & \ddots & 0       & \vdots\\
\vdots          &\ldots& 0  & 1 & -T^{-1}     & 0  \\
0               &\ldots& \ldots&0 & 1     & -T^{-1}   \\
-T^{-1} (P_+ -m P_-) & 0    & \ldots&\ldots   & 0      & 1    
\end{array}
\right]\,,
\end{eqnarray}
for which we can perform a UDL decomposition around the top left block:
\begin{equation}
\left(
\begin{array}{cc}
D & C \\
B & A
\end{array}
\right)
= 
\left(
\begin{array}{cc}
1  &  C A^{-1} \\
0  & 1
\end{array}
\right)
\left(
\begin{array}{cc}
S_\chi & 0\\
0 & A
\end{array}
\right)
\left(
\begin{array}{cc}
1 &  0 \\
A^{-1} B  & 1
\end{array}
\right)\,.
\end{equation}
Denoting the left and right factors as $U$ and $L(m)$ respectively, we 
write this factorisation as $D_\chi^5 = U D_S(m) L(m) $.
The determinants of the $U$ and $L(m)$ are unity, and the determinant of the product is
$
\det D_\chi^5 = \det A \det S_\chi = \det S_\chi,
$ where
\begin{eqnarray}
S_\chi(m)&=& -(1 + T^{-L_s}) \gamma_5 
\left[ \frac{1+m}{2} + \frac{1-m}{2}\gamma_5 \frac{T^{-L_s}-1}{ T^{-L_s}+1} \right].
\end{eqnarray}
We can see that after the removal of the determinant of the Pauli Villars fields  we are left with the 
determinant of an effective overlap operator, which is the following rational function of the kernel:
\begin{equation}
\det D_{PV}^{-1} D(m)= \det D_{ov}  = \det\left(\frac{1+m}{2} + \frac{1-m}{2}\gamma_5 
\frac{ (1+ H_M)^{L_s}-(1 - H_M)^{L_s}}
     { (1+ H_M)^{L_s}+(1 - H_M)^{L_s}}\right)\,.
\end{equation}
We identify $D_{ov}$ as an approximation to the overlap operator with 
approximate sign function
\begin{equation}
\epsilon(H_M) = \frac{ (1+ H_M)^{L_s}-(1 - H_M)^{L_s}}
                     { (1+ H_M)^{L_s}+(1 - H_M)^{L_s}}\,,
\label{eq:sign_approx}
\end{equation}

The approximate overlap operator can be written as
$
D_{ov} = S_\chi(m=1)^{-1} S_\chi(m) .
$
If we solve the following 5d system of equations and substitute the UDL decomposition,
\begin{eqnarray}
D_\chi^5(m=1)^{-1} D_\chi^5(m) \phi &=& 
\left(
q,
0,
\ldots
0
\right)^T,
\end{eqnarray}
the approximate overlap operator can be expressed
in terms of the $\bar{\psi}$ basis fields, 
\begin{eqnarray}
D_{ov} &=& S_\chi(m=1)^{-1} S_\chi(m)  
      = \left[{\cal P}^{-1} D^5_{GDW}(m=1)^{-1}  D^5_{GDW}(m) {\cal P}\right]_{11}\,.
\end{eqnarray}
The cancellation the  Pauli-Villars term can be expressed 
in terms of \emph{unmodified} generalised domain wall matrix $D^5_{GDW}$. 
The contact term can be subtracted from the overlap propagator. We define
\begin{eqnarray}
\tilde{D}_{ov}^{-1}&=&\frac{1}{1-m}\left[{D}_{ov}^{-1} - 1\right]
 =\frac{1}{1-m} 
\left\{ {\cal P}^{-1} D^5_{GDW}(m)^{-1}  
\left[ D^5_{GDW}(m=1)  -        D^5_{GDW}(m) \right]{\cal P}
\right\}_{11}\,.
\end{eqnarray}

Now, the difference 
$\left[ D^5_{GDW}(m=1)  -        D^5_{GDW}(m) \right]_{ij} = 
(1-m)\left[ P_- \delta_{i,L_s}\delta_{j1} 
+ P_+ \delta_{i,1}\delta_{j,L_s}\right]$.
This relation is simpler to interpret in our convention than with the
convention from \cite{Brower:2012vk}: the mass term is applied to our five dimensional
surface fields without field rotation. With this,
\begin{eqnarray}
\tilde{D}_{ov}^{-1}&=&
\left\{ {\cal P}^{-1} D^5_{GDW}(m)^{-1}  R_5 {\cal P}
\right\}_{11}\,.
\end{eqnarray}
This is just the normal valence propagator of the physical DWF fields 
$q = ({\cal P}^{-1} \psi)_1$
and $\bar{q} = (\bar{\psi} R_5 {\cal P})_1$.
We see that the usual domain wall valence propagator has always contained both the contact term
subtraction and the appropriate multiplicative renormalisation of the overlap
fermion propagator. As a result, the issues of lattice artefacts in NPR
raised in reference~\cite{Maillart:2008pv} have never been present in domain wall valence analyses.
This was guaranteed to be the case because Shamir's 5d construction is designed to
exactly suppress chiral symmetry breaking in the limit of infinite $L_s$, including
any contact term. 
For later use, we may also consider the propagator into the bulk from a surface field $q$ for Mobius fermions
\begin{eqnarray}
\langle Q_s \bar q \rangle &=& 
\left[ 
{\cal P}^{-1} D^5_{GDW}(m)^{-1} R_5 {\cal P} 
\right]_{s1}
=
\frac{1}{1-m} 
\left\{
L^{-1}(m) 
\left(
\begin{array}{c|c}
S_\chi^{-1}(m) S_\chi(1) & 0 \\
\hline
0 & \ident
\end{array}
\right)
L(1) -\ident
\right\}_{s1}.
\end{eqnarray}
Now, 
\begin{equation}
\begin{array}{ccc}
L(m) = 
\left(
\begin{array}{c|c}
1 & 0 \\
\hline
\begin{array}{c}
-T^{-(L_s-1)} (P_+-m P_-)\\
\vdots \\
-T^{-1} (P_+-m P_-)
\end{array} & \ident
\end{array}
\right)
&\quad;\quad&
L(m)^{-1} = 
\left(
\begin{array}{c|c}
1 & 0 \\
\hline
\begin{array}{c}
T^{-(L_s-1)} (P_+-m P_-)\\
\vdots\\
T^{-1} (P_+-m P_-)
\end{array} & \ident
\end{array}
\right)
\end{array}
\end{equation}
Finally, applying ${\cal {P}}$, we have the five dimensional propagator from a physical field,
\begin{eqnarray}
G_q  = {\cal P} \langle Q_s \bar q\rangle
&=& [P_+  +P_- T^{-1}] 
\left(
\begin{array}{c}
T^{-(L_s-1)}   \\
T^{-(L_s-2)}  \\
\vdots\\
T^{-1} \\
1
\end{array}
\right)
[1+T^{-L_s}]^{-1} D_{ov}^{-1}.
\label{BulkPropagator}
\end{eqnarray} 

The connection between domain wall systems and the overlap, well established in the literature
and reproduced in this section, is needed in the following derivation of conserved currents. 

\section{Conserved vector and axial currents}
\label{appendix-mobiusconservedcurrents}

The standard derivation of lattice Ward identities proceeds
as follows. A change variables of the fermion fields $\psi$ and $\bar \psi$ 
at a single site $y$ is performed:
$
    \psi^\prime_y = \psi_y - i \alpha \psi_y$ ; 
$\bar \psi^\prime_y = \bar \psi_y + i \bar \psi_y \alpha$
under the path integral, the Jacobian is unity, 
and the partition function is left invariant
\begin{eqnarray}
Z^\prime
&=&
  \int d\bar\psi d\psi e^{-S[\bar\psi,\psi]} 
    \left\{1 - i \alpha \left[ \frac{\delta S}{\delta \psi_y} \psi_y
                     - \bar \psi_y \frac{\delta S}{\delta \bar\psi_y} 
	\right] \right\}
= Z.
\end{eqnarray}
Hence,
\begin{equation}
 \langle \frac{\delta S}{\delta \psi_y}\psi_y - \bar \psi_y \frac{\delta S}{\delta \bar\psi_y} 
\rangle = 0.
\end{equation}
The Wilson action gives eight terms from varying $\bar{\psi}_y$ and
eight terms from varying $\psi_y$:
\begin{eqnarray}
\bar\psi \delta_y (D_W) \psi = \Delta^-_\mu J^W_\mu (y)
&=&
\label{WilsonDivergence}
\Delta^-_\mu \left[
 \bar\psi_y \frac{1-\gamma_\mu}{2} U_\mu(y) \psi_{y+\hat\mu}
-\bar\psi_{y+\hat\mu} U^\dagger_\mu(y) \frac{1+\gamma_\mu}{2} \psi_y
\right]= 0,
\end{eqnarray}
where $\Delta^-_\mu$ is the backwards discretized derivative.
An equivalent alternate approach may be taken, however, and this is a better
way to approach non-local actions such as the chiral fermions. 
Gauge symmetry leaves the action invariant at $O(\alpha)$ under the simultaneous active substitution,
\begin{equation}
U_\mu(y) \to (1+i\alpha) U_\mu(y) 
 ; \quad
U_\mu(y-\hat\mu) \to  U_\mu(y-\hat\mu) (1-i\alpha) 
 ; \quad
     \psi_y \to (1+ i \alpha )\psi_y 
 ; \quad
\bar \psi_y \to \bar  \psi_y (1 - i \alpha)
.\end{equation}
A change variables on the fermion fields at site $y$
may be performed simultaneously to absorb the phase on the fermions,
$
\psi^{\prime}_y =  (1+ i \alpha) \psi_y$; 
$
\bar \psi^{\prime}_y = \bar \psi_y (1 - i \alpha).
$
Under the path integral, the Jacobian is unity, and the phase associated with the fermion is absorbed.
We can now view the change in action as being associated with the \emph{unabsorbed}
phases on the eight gauge links connected to site $y$. 
\begin{equation}
Z^\prime = Z =
 \int d\bar\psi^\prime d\psi^\prime e^{-S[\bar\psi^\prime,\psi^\prime,U]}
    \left\{1 + i \alpha \sum_\mu \left[\frac{\delta S}{\delta U_\mu(y)^{ij}} U_\mu(y)^{ij}
    - \frac{\delta S}{\delta U_\mu(y-\mu)^{ij}} U_\mu(y-\mu)^{ij}
	\right] \right\}.
\end{equation}
For a gauge invariant Lagrangian we can \emph{always} 
use a picture where the same change in action, and same current conservation law 
may be arrived at by differentiating with respect to the eight links connected to a site
\begin{equation}
\langle \sum_\mu \left[
 \frac{\delta S}{\delta U_\mu(y)^{ij}} U_\mu(y)^{ij}
    -                 
 \frac{\delta S}{\delta U_\mu(y-\mu)^{ij}} U_\mu(y-\mu)^{ij}
	\right]\rangle = 0.
\end{equation}

This arises because the phase freedom of fermions and of gauge fields are necessarily coupled and inseparable in a gauge theory.
For the nearest neighbour Wilson action this generates the same eight terms entering $\Delta^-_\mu J_\mu =0$.
In the case of non-local actions, the Dirac matrix, whatever it is, can be viewed as a sum of gauge covariant paths. 
When we generating a current conservation law from $U(1)$ rotation of the fermion field
at site $y$, we sum over all fields $\bar \psi(x)$ and $\psi(x)$ connecting through the Dirac matrix $D(x,y)$ to the fixed site
$\psi(y)$ and $\bar\psi(y)$. The following sum is always constrained to be zero for all $y$, and is identical to that 
found by Kikukawa and Yamada\cite{Kikukawa:1998py}:
\begin{equation}
\sum_x \bar{\psi}_x D(x,y) \psi_y - \bar{\psi}_y D(y,x) \psi_x = 0.
\label{sum_df}
\end{equation}
The partitioning of this sum of terms, into a paired \emph{discrete divergence operator} and \emph{current} is not obvious,
and it is cumbersome to generate Kikukawa and Yamada's \emph{non-local kernel}.
We may derive the same sum of terms by differentiating with respect to the
8 links connected to site $y$.
\begin{equation}
\langle
\sum_\mu \left[
\begin{array}{c}
 \frac{\delta S}{\delta U_\mu(y)^{ij}} U_\mu(y)^{ij} 
    -                 
 \frac{\delta S}{\delta U_\mu(y-\mu)^{ij}} U_\mu(y-\mu)^{ij}
\end{array}
	\right]
\label{sum_du}
\rangle = 0
\end{equation}
The structure of eqn.~\ref{sum_du} \emph{always} lends itself interpretation as a backwards finite difference.
For a non-local action the differentiation eqn.~\ref{sum_du} appears to generate a lot more terms than the fermion field differentiation
eqn.~\ref{sum_df}. The reason is clear: these extra terms are constrained by gauge symmetry to sum to zero, but only after
cancellation between the different terms in eqn.~\ref{sum_du}. 
Specifically, we consider an action constructed as the product of Wilson matrices:
\begin{equation}
S = \sum_{xyzw} \bar \psi_x D_W(x,y) D_W(y,z)D_W(z,w) \psi(w).
\end{equation}

The link variation approach gives three terms, each of which are conserved under a
nearest neigbour difference divergence: varying with respect to the 8 links we obtain
via the product rule

\begin{equation}
\delta_y (\bar\psi D_W D_W D_W \psi) \psi = \psi \left[ (\delta_y D_W) D_W D_W + D_W (\delta_y D_W)D_W  +D_W D_W(\delta_y D_W) \right] \psi 
\end{equation}

Each of these contributions contain a backwards difference operator and it is trivial
to split this into a divergence and corresponding conserved current using eqn.~(\ref{WilsonDivergence}).
The above comment is generally applicable to any function of the Wilson matrix.
We take this approach to establish the exactly conserved vector current of an approximate overlap operator, where the
approximation is represented by a rational function. 
We will also establish that matrix elements of this current are identical to those of the Furman and Shamir
approach~\cite{Furman:1994ky} in the case of domain wall fermions.
The Furman and Shamir approach will then be used
to also establish an axial Ward identity  for our generalised M\"obius domain wall fermions.
 under which an explicitly known defect arises.
This is important in both renormalising lattice operators and also in determining the most appropriate measure
of residual chiral symmetry breaking in our simulations.
We construct the conserved vector current by determining
the variation in the overlap Dirac operator $\delta_y D_{ov}$
\begin{eqnarray}
\delta_y D_{ov} &=& \frac{1-m}{2} \gamma_5 \left\{
\delta_y (\frac{1}{1+T^{-Ls}})  [1-T^{-Ls}]
+\frac{1}{1+T^{-Ls}} \delta_y (1-T^{-Ls})
\right\}\nonumber
= (1-m)\gamma_5 \delta_y \left(\frac{1}{1+T^{-Ls}}\right).
\end{eqnarray}

We can similarly find the variation in $T^{-1}$ induced by a variation in $D_W$, 
where the variation in $D_W$ is just the backwards divergence of the standard Wilson conserved 
current operator. Denoting,
\begin{eqnarray}
T^{-1} &=& -(\tilde{Q}_-)^{-1}\tilde{Q}_+\nonumber\\
\tilde{Q}_- &=&  D_+^s P_- - D_- P_+ = D_- \gamma_5 Q_-  \nonumber\\
\tilde{Q}_+ &=&  D_+^s P_+ - D_- P_- = D_- \gamma_5 Q_+  ,
\end{eqnarray}
we see that 
\begin{eqnarray}
\delta_y(T^{-1})  
 &=&  -\tilde{Q}_-^{-1} \delta_y(D_W) \left\{  (b P_-  + c P_+ ) T^{-1} + b P_+ +c P_- \right\}. 
\end{eqnarray}
Since
\begin{eqnarray}
\begin{array}{ccccccc}
\tilde{Q}_- P_- &=& (1+b D_W)P_- 
&;&
\tilde{Q}_+ P_- &=& (c D_W-1)P_- \nonumber\\
\tilde{Q}_- P_+ &=& (c D_W-1)P_+ 
&;&
\tilde{Q}_+ P_+ &=& (1+b D_W)P_+ ,
\end{array}
\end{eqnarray}
we may rexpress the identity 
\begin{eqnarray}
\tilde{Q}_-^{-1}(P_+ + P_-) &=& \frac{\tilde{Q}_-^{-1}}{b+c} \left[ \tilde{Q}_+ (c P_+ - b P_- ) + \tilde{Q}_- (c P_- - b P_+) \right],
\end{eqnarray}
and this lets us find a symmmetrical form:
\begin{eqnarray}
(b+c)\delta_y(T^{-1}) &=& 
\left[ b [P_+ - T^{-1}P_-] + c [T^{-1} P_+ - P_- ]  \right]
\delta_y(D_W) 
\left[  b [ P_+ +  P_- T^{-1}] + c [ P_+ T^{-1} + P_- ]\right]. \label{deltaTinv}\nonumber
\end{eqnarray}
We may now look at the variation of the term $T^{-L_s}$
\begin{equation}
\delta_y(T^{-L_s}) =
\sum\limits_{s=1}^{L_s}
T^{-(s-1)} 
\left[ 
\begin{array}{c} 
b [P_+ - T^{-1}P_-] \\+ c [T^{-1} P_+ - P_- ]
\end{array}  \right]
\delta_y(D_W) 
\left[
\begin{array}{c} 
  b [ P_+ +  P_- T^{-1}] \\+ c [ P_+ T^{-1} + P_- ]
\end{array}
\right]
T^{- (L_s-s)}. 
\end{equation}
Pulling these results together, we find
\begin{equation}
\delta_y D_{ov} =  -\frac{1-m}{b+c} \gamma_5 \frac{1}{1+T^{-Ls}}
\left( \sum\limits_{s=1}^{L_s}
T^{-(s-1)} \delta_y(T^{-1}) T^{- (L_s-s)} 
\right)
\frac{1}{1+T^{-Ls}}.
\end{equation}

The terms may be expanded until insertions of the
the backwards divergence of the Wilson current are reached, eqn.~\ref{WilsonDivergence}.
Gauge symmetry then implies the conservation of the obvious current and 
the vector Ward identities can be constructed. For example, we may take
as source $\eta^{j j^\prime\alpha\alpha^\prime} (z) = \delta_{jj\prime}\delta_{\alpha\alpha^\prime} \delta^4(z-x)$
and a two point function of the conserved current may be constructed as

\begin{equation}
\label{MobiusMatrixElement}
\begin{array}{c}
\Delta^-_\mu \langle \bar \psi \gamma_\nu \psi(x)| {\cal V}_\mu(y) \rangle =
{\rm Tr} \gamma_\nu \gamma_5 \eta^\dagger D_{ov}^{-\dagger} \gamma [1+T^{-L_s}]^{-1}
\left\{\sum\limits_{s=0}^{L_s-1}
T^{-s} \delta_y(T^{-1}) T^{-(L_s-1-s)}
\right\}
[1+T^{-L_s}]^{-1} D_{ov}^{-1}\eta
\end{array}.
\end{equation}

Note that  when $c=0$ the insertion of eqn.~\ref{deltaTinv} contains only terms such as 
$
[ P_-  T^{-1} + P_+ ],
$
which are also present in the surface to bulk propagator eqn.~\ref{BulkPropagator}.
As one would expect, when we take $b$ and $c$ to represent domain wall fermions, 
the two point function of our exactly conserved vector current - derived from the four dimensional effective 
action - exactly matches the matrix element of the vector current
constructed by Furman and Shamir~\cite{Furman:1994ky}, eqn.~(2.21), from a five dimensional interpretation
of the action.
Since the Furman and Shamir current was easily constructed from the five dimensional propagator 
eqn.~(\ref{BulkPropagator}) one might hope to do the same in the generalised approach to domain wall fermions.
To play a similar trick for the $c$ term we would need to generate the terms
$
P_- [1+T^{-L_s}]^{-1} D_{ov}^{-1}$,
and $
 P_+  T_1^{-1}[1+T^{-L_s}]^{-1} D_{ov}^{-1}.$
These are not manifestly present in eqn.~\ref{bulkpropagatormob}.
However, the presence of the contect term on the $s=0$ slice can be removed after a propagator
calculation. We define this slice as 
$
S(x) = \langle Q_0 \bar q\rangle = \frac{1}{1-m}\left( D_{ov}^{-1}(m)  - \ident\right).
$
In a practical calculation the source vector $\eta$ may be used to eliminate the contact term by forming
\begin{equation}
(1-m)S(x)\eta + \eta = D_{ov}^{-1}(m) \eta =  [1+T^{-L_s}] [1+T^{-L_s}]^{-1} D_{ov}^{-1} \eta.
\end{equation}
By applying $P_+$ and $P_-$ we find we have the following set of vectors
\begin{equation}
\left(
\begin{array}{c}
P_+ ,
P_- T^{-L_s},
P_+[1+T^{-L_s}],
P_-[1+T^{-L_s}]
\end{array}
\right)^T[1+T^{-L_s}]^{-1} D_{ov}^{-1},
\end{equation}
and we may eliminate to form a $L_s+1$ vectors from a 4d source $\eta$
\begin{equation}
T(s) = 
\left(
\begin{array}{c}
1   ,
T^{-1},
\ldots,
T^{-L_s}
\end{array}
\right)^T \left[ 1+ T_1^{-1}\cdots T_{L_s}^{-1}\right]^{-1} D_{ov}^{-1}(m)\eta.
\end{equation}
This may be used to construct
\begin{equation}
\left[  b [ P_+ +  P_- T^{-1}] + c [ P_+ T^{-1} + P_- ]\right] T^{s},
\end{equation}
for $s\in \{ 0 \ldots L_s-1 \} $, and by contracting these vectors through the Wilson conserved
current the the matrix element eqn.~\ref{MobiusMatrixElement} can be formed a \emph{very} similar manner to the
standard DWF conserved vector current. When $c = 0$ the matrix element reduces to being \emph{identical}
to that for the Furman and Shamir vector current.
A flavour non-singlet axial current, almost conserved under a backwards difference operator, can now also be constructed following Furman and Shamir.
We associate a fermion field rotation 
\begin{equation}
\begin{array}{ccc}
\psi(x,s) \to \left\{ 
\begin{array}{ccc}
e^{i \alpha \Gamma(s)} \psi(x,s) & ; & x=x_0 \\
 \psi(x,s) &;& x\ne x_0
\end{array}
\right.
&;&
\Gamma(s) \to \left\{ 
\begin{array}{ccc}
-1 &;& 0\le s < L_s/2\\
1 &;& L_s/2 \le s
\end{array}
\right. .
\end{array}
\end{equation}
We acquire a related (almost) conserved axial current, whose pseudoscalar matrix element is
\begin{equation}
\label{WTI}
\begin{array}{c}
\Delta^-_\mu \langle \bar \psi \gamma_5 \psi(x)| {\cal A}_\mu(y) \rangle =
{\rm Tr}  [\eta^\dagger \tilde D_{ov}^{-\dagger} \gamma_5 ]
[1+T^{-L_s}]^{-1}
\left\{
\sum\limits_{s=0}^{L_s-1}T^{-s} \Gamma(s) \delta_y(T^{-1}) T^{-(L_s-1-s)}
\right\}
[1+T^{-L_s}]^{-1}
 D_{ov}^{-1} \eta
\end{array}
\end{equation}
This generalisation of the Furman and Shamir approach 
induces the same $J_{5q}$ midpoint density defect that
arose for DWF, and the axial Ward identity is 
\begin{equation}
\label{mobpcac}
\Delta^-_\mu \langle \bar \psi \gamma_5 \psi(x)| {\cal A}_\mu(y) \rangle = \langle \bar \psi \gamma_5 \psi(x)|2m P(y) +2J_{5q}(y) \rangle .
\end{equation}
This allows us to retain the usual definition of the residual mass in the case of M\"{o}bius domain wall fermions.
We emphasize that the definition,
$$m_{res} =\left. \frac{\langle \pi(\vec{p}=0) | J_{5q}\rangle}
{ \langle \pi(\vec{p}=0) | P \rangle}\right|_{m=-m_{res}},$$
via the zero-momentum pion matrix element of $J_{5q}$ is important, because
the PCAC relation, $$\langle \pi(\vec{p}=0) |2m P +2J_{5q} \rangle = 0,$$  
guarantees that the low momentum lattice pions are massless. This is the appropriate measure of chiral
symmetry breaking for the analysis of the chiral expansion.

\begin{figure}[hbt]
\begin{center}
\includegraphics[width=0.6\textwidth]{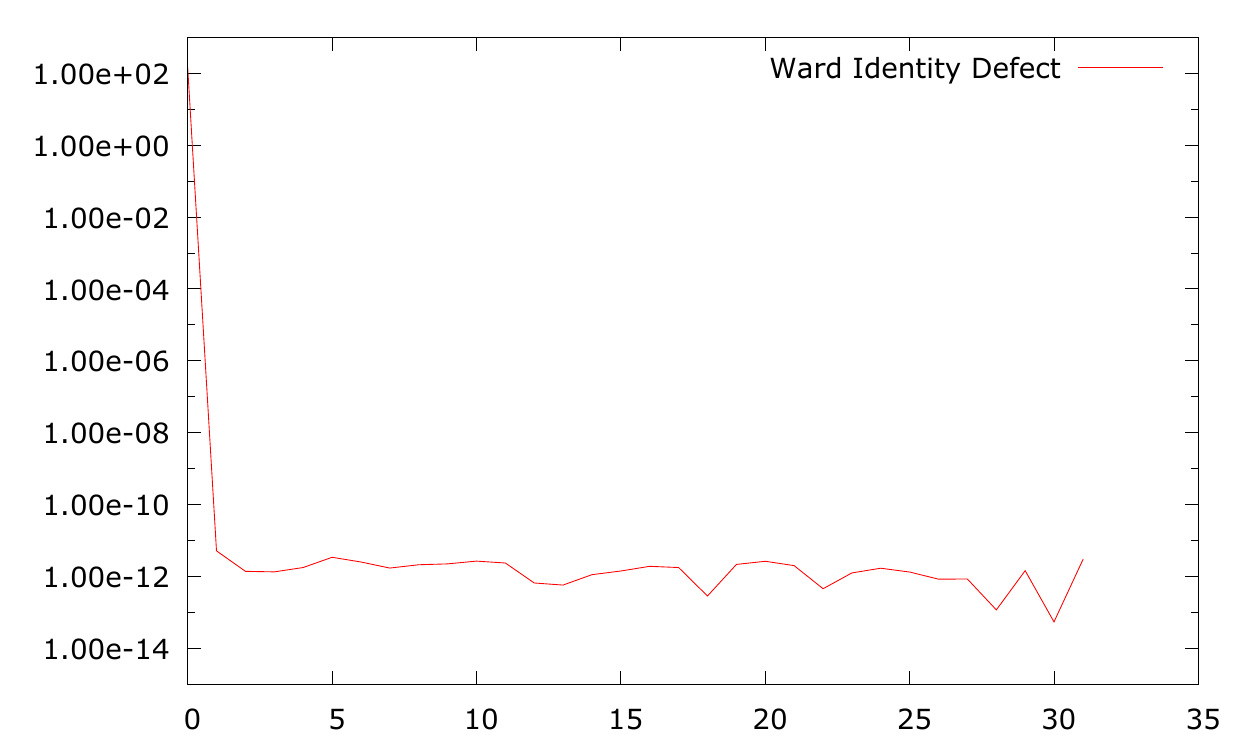}
\end{center}
\caption{
As a numerical proof of the correctness we display the 
difference of the left and right hand sides of  the axial Ward identity
evaluated on a  $16^3$ configuration with a point source. The defect is of order the convergence error.
}
\end{figure}

\section{Acknowledgements}
PB wishes to thank Norman Christ, Antonin Portelli, Shoji Hashimoto, Taku Izubuchi, Richard Browerand his colleagues
in the RBC-UKQCD collaboration for useful discussions.

\end{document}